\documentclass[12pt]{JHEP3}
\usepackage{epsfig}
\usepackage{amsmath}
\usepackage{amssymb}

\newcommand{\p}{\partial}
\newcommand{\Tr}{{\rm Tr}}

\title{
Higher Dimensional Recombination of Intersecting D-branes}

\author{
Satoshi Nagaoka\\
Institute of Physics, University of Tokyo\\
Komaba, Tokyo 153-8902, Japan\\
E-mail: \email{nagaoka@hep1.c.u-tokyo.ac.jp}}

\abstract{
We study recombinations of D-brane systems 
intersecting at more than one angle using super Yang-Mills theory.
We find the condensation of an off-diagonal tachyon mode relates to the 
recombination, as was clarified for branes
at one angle in hep-th/0303204.
For branes at two angles,
after the tachyon mode between two D2-branes condensed, 
D2-brane charge is distributed in the bulk near the intersection point.
We also find that,
when two intersection angles are equal, the off-diagonal lowest mode  
is massless, and a new stable non-abelian configuration, 
which is supersymmetric
up to a quadratic order in the fluctuations, is obtained by the  
deformation by this mode.}


\preprint{UT-Komaba/03-19, hep-th/0312010}

\begin{document}

\section{Introduction}

Low energy dynamics of D-brane is well known by the analysis 
using super Yang-Mills theories \cite{Wit}.
One of the D-brane systems which can be studied in Yang-Mills theories
is intersecting D-brane systems.
Fluctuation spectra of intersecting D-branes on a torus are
studied by using Yang-Mills action in \cite{HasTay}.
In intersecting brane systems, recombination is important 
in a context of string phenomenology.
Standard Model Higgs mechanism is realized as a brane 
recombination \cite{higgs}. On the other hand,
it plays an important role to realize an inflation 
in brane world scenario \cite{cosmology}.
In Yang-Mills theories, the recombination is studied in
\cite{Recombi,Morosov} for branes at one angle\footnote{It
is also discussed in tachyon field theory \cite{Huang1} and 
Matrix theory \cite{Huang2}. Intersecting D-branes with a separation 
are discussed in \cite{Sato}.}.
Fundamental strings stretched between intersecting D-branes are also
studied in Yang-Mills theories in \cite{HT}.
We extend the study of recombination of intersecting D-branes 
to the case with more than one intersection angle.
This enables us to study
more complicated intersecting brane systems, 
which may appear in the context of Standard Model realized by
 intersecting D-branes.
To clarify the mechanism of recombination in higher dimensions 
 is one of the aims of this paper.

When we consider two intersection angles,
we find a supersymmetric configuration where two intersection angles are
equal.
It is known that supersymmetric intersecting branes are embedded 
into spacetime on a calibrated surface 
to minimize their worldvolume \footnote{For a recent review, 
see \cite{Smith}.}.
Calibration equations are realized as BPS conditions in abelian 
Dirac-Born-Infeld actions \cite{GP}.
An embedding realized by the calibration geometry
is called in \cite{CL} as an abelian embedding.
The dynamics of multiple D-branes is described by non-abelian
Born-Infeld(NBI) actions. In the non-abelian cases, there are other embeddings 
which cannot be realized by the calibration geometry, and further, which
are constructed from the components including off-diagonal 
elements of the fields in the NBI actions.
We call such embeddings as non-abelian embeddings.
There are various kinds of the non-abelian embeddings and they include 
many interesting characters
but it has not yet been considered except for a few cases \cite{CL}.
It is difficult to know the full NBI action because 
in the non-abelian cases, slowly varying field approximation
is meaningless \cite{Bilal} 
and we must consider derivative terms together with field strengths 
\footnote{We know up to and including $F^6$ terms of the NBI action now 
\cite{F^6action,Nagaoka1}.}.
But there are some non-abelian embeddings which Yang-Mills 
analysis is valid for.
We believe that the results of such Yang-Mills analysis can be 
lifted smoothly to full NBI analysis.

In this paper, we study recombination of 
D2-branes intersecting at two angles.
When two angles are not equal, 
the lowest mode of a Neveu-Schwarz sector
is tachyonic, and considering a condensation of this mode,
we obtain deformed intersecting branes.
For branes at one angle, the recombination occurs locally near the 
intersection point and we can
describe this phenomenon by a condensation of a tachyon mode
which is localized at an intersection point \cite{Recombi}.
The final state in this decay process is a set of two parallel D2-branes.
For branes at two angles,
the final state after the recombination is expected 
as a brane configuration which preserves $1/4$ of the 
supersymmetries \cite{Uranga}.
The tachyon mode we consider here is localized 
at the intersection point even in this case, 
therefore, the condensation of the tachyon mode describes 
the first step of the recombination, plays a role of a trigger.
After the tachyon mode has condensed,
the branes can not be realized as a simple geometrical surface, because
two transverse scalar fields can not be diagonalized 
by any gauge transformation simultaneously. 
We find D2-brane charge distributed in the bulk
near the intersection point after the tachyon condensation.
When two intersection angles are equal, 
we obtain a massless mode which appears in an off-diagonal spectrum 
in Yang-Mills theory.
By considering a deformation by this massless mode, 
we obtain a configuration which is supersymmetric up to a quadratic order
in the fluctuations. In this configuration,
D2-brane charge is again distributed in the bulk
near the intersection point.
The massless deformation considered here does not seem to be expressed by
 any abelian calibration geometry\footnote{Calibration of
supersymmetric intersecting D-branes in terms of world-volume field
theory is discussed in \cite{EGHK}.
} and this is an interesting example 
of the non-abelian embeddings.

In section 2, 
we discuss a recombination of D2-branes intersecting at two angles.
When two intersection angles are equal, the brane configuration is 
supersymmetric and there is an off-diagonal massless mode
which is discussed in section 3. 
The connections with higher orders in field strengths
are discussed in section 4. Section 5 is devoted to a conclusion and
discussions.

\section{Recombination of D-branes intersecting at two angles}

Before performing Yang-Mills analysis, we see
a mass spectrum of a fundamental string stretched between two 
D-branes intersecting at multiple angles $\theta_i(i=1,\cdots,a)$, 
which is obtained in the worldsheet analysis as \cite{BDL,Jab}
\footnote{See also \cite{Ohta}.}
\begin{align} \label{stringmass}
m^2_j=\frac{1}{2\pi\alpha'} \sum_{i=1}^a (2 n_i-1) \theta_i \pm 2\theta_j
\quad , 
\end{align}
where $n_i \in \mathbb{N}$.

\subsection{Fluctuation modes}

We start with a worldvolume effective action for two D2-branes,
which is obtained by the dimensional reduction of 
a (9+1) dimensional SU(2) Yang-Mills action,
\begin{align} \label{effac}
S= -T \ \Tr \int d^2x dt \left[ (D_a Y^i)^2 +\frac{1}{2}F_{ab}^2 
-\frac{1}{2} [Y^i,Y^j]^2 \right] \ .
\end{align}
The indices $a,b=0,1,2$ denote directions along the 
worldvolume, and $i,j=3,\cdots,9$
denote directions of transverse collective coordinates. 
$F_{ab}$ and $D_a Y^i$ are defined by
\begin{align}\notag
&F_{ab}=\p_a A_b -\p_b A_a -i [A_a,A_b] \ , \\
&D_a Y^i=\p_a Y^i - i [A_a,Y^i] \ ,
\end{align}
where $A_a$'s are worldvolume gauge fields and $Y^i$'s are transverse
scalar fields. Let us consider an intersecting D2-brane system.
It is sufficient for us to consider the situation that 
the two D2-branes are embedded in 4 dimensions and 
do not extend in other dimensions.
We take coordinates of the embedding space as $1,2,8$ and $9$ and
others as $3,\cdots,7$.
We describe intersection angles
as $\theta_1$ in $x_1$-$Y^9$ plane and $\theta_2$ in $x_2$-$Y^8$ plane
here.
We consider a classical solution representing the intersecting D-branes
\begin{align} \label{sol1}
Y^9=q_1 x_1 \sigma^3 ,
\quad
Y^8=q_2 x_2 \sigma^3 ,
\quad
A_a=0 \ ,
\end{align}
where $q_\alpha (\alpha=1,2)$ is related to the intersection angle
$\theta_\alpha$ as 
$\theta_\alpha \equiv 2 \tan^{-1} (2 \pi \alpha' q_\alpha)$
and we assume $q_\alpha >0$ here.

Let us consider off-diagonal parts of fluctuations around the solution
\begin{align} \label{flu1} \notag
&Y^9=q_1 x_1 \sigma^3 +f_1(x_a) \sigma^1-\bar{f}_1 (x_a)\sigma^2 , \quad
Y^8=q_2 x_2 \sigma^3 +f_2(x_a) \sigma^1 -\bar{f}_2 (x_a) \sigma^2 \ , \\
&A_1=g_1(x_a) \sigma^1-\bar{g}_1(x_a)\sigma^2 , \quad
A_2=g_2(x_a) \sigma^1-\bar{g}_2(x_a)\sigma^2 . 
\end{align}
We adopt a gauge condition $A_0=0$.
Diagonal parts of the fluctuations are decoupled from the off-diagonal
fluctuations at the quadratic order, therefore we neglect the former ones.

\begin{figure}[t]
\begin{center}
\begin{minipage}{15cm}
\begin{center}
\includegraphics[width=14cm]{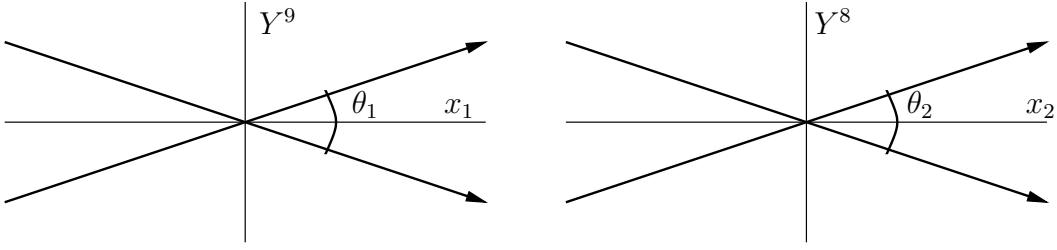} 
\put(-230,50){$x_1$}
\put(-10,50){$x_2$} 
\put(-90,80){$Y^8$} 
\put(-300,80){$Y^9$}
\put(-265,50){$\theta_1$}
\put(-55,50){$\theta_2$}
\caption{Intersecting D2-branes} 
\end{center}
\end{minipage}
\end{center}
\end{figure}

The Lagrangian quadratic in the fluctuations is calculated as
\begin{align}\notag
L=\sum_{i,j=1,2}&
\Big(-(\p_0 f_i)^2 -
4q_if_i\bar{g}_i+
(\p_i f_j+2\bar{g}_iq_jx_j)^2
-(\p_0 \bar{g}_i)^2
+(\p_i\bar{g}_j-\p_j\bar{g}_i)^2 \\ 
&+4 (q_i x_i f_j-q_j x_j f_i)^2 
+(f_\alpha \to \bar{f},\bar{g}_\alpha \to g ) \Big)
\ .
\end{align}
The combinations ($f_\alpha , \bar{g}_\alpha $) and 
($\bar{f}_\alpha , g_\alpha $) are decoupled each other at this order,
therefore we neglect the ($\bar{f}_\alpha , g_\alpha$)
pair. From now on, we denote $\bar{g}$ as $g$ for simplicity.
The equations of motion for the fluctuations are written as
 \begin{align}
4 \left(
\begin{array}{cc}
O_{11}&O_{12}\\
O_{21}&O_{22}
\end{array}\right)
\left(
\begin{array}{c}
\vec{V}_1 \\ \vec{V}_2 
\end{array}\right)
=0 \ ,
\end{align}
where 
\begin{align}\notag
&O_{11}=
\left(
\begin{array}{cc}
\p_0^2 -\p_1^2-\p_2^2+4 q_2^2 x_2^2& -4q_1-2q_1x_1 \p_1 
\\ -2q_1+2q_1x_1 \p_1& 4q_1^2x_1^2+4q_2^2x_2^2+
\p_0^2-\p_2^2
\end{array}\right) , \\ \notag
&O_{12}=O_{21}=
\left(
\begin{array}{cc}
-4 q_1 q_2 x_1 x_2&-2q_1x_1\p_2
\\2q_2x_2\p_1&\p_1 \p_2
\end{array}\right) , \\
&O_{22}=
\left(
\begin{array}{cc}
\p_0^2-\p_1^2-\p_2^2+4 q_1^2 x_1^2&-4q_2-2q_2x_2\p_2
\\ -2q_2+2q_2x_2\p_2
&4q_1^2x_1^2+4q_2^2x_2^2+\p_0^2-\p_1^2
\end{array}\right) \ ,
\end{align}
and 
\begin{align}
\vec{V}_1=\left(
\begin{array}{c}
f_1(x_a)\\g_1(x_a)
\end{array}\right), \quad
\vec{V}_2=\left(
\begin{array}{c}
f_2(x_a)\\g_2(x_a)
\end{array}\right) \ .
\end{align}
Expanding the fluctuations by the mass eigenfunctions as
\begin{align}
\vec{V}_1(x_1,x_2,t)=\sum_{n\geq 0} 
\left(
\begin{array}{c}
\tilde{f}_{1n}(x_1,x_2)\\ \tilde{g}_{1n}(x_1,x_2)
\end{array}\right)
C_{1n} (t)\ , \quad
\vec{V}_2(x_1,x_2,t)=\sum_{n\geq 0} 
\left(
\begin{array}{c}
\tilde{f}_{2n}(x_1,x_2)\\ \tilde{g}_{2n}(x_1,x_2)
\end{array}\right)
C_{2n} (t) \ ,
\end{align}
where $C_{in}$ $(i=1,2)$'s satisfy the following equations 
\begin{align}
(\p_0^2+m_{in}^2) C_{in}(t)=0 \ ,
\end{align}
we obtain the differential equations written as
 \begin{align}
4 \left(
\begin{array}{cc}
O'_{11}&O_{12}\\
O_{21}&O'_{22}
\end{array}\right)
\left(
\begin{array}{c}
\vec{V}'_{1n} \\ \vec{V}'_{2n} 
\end{array}\right)
=\left(
\begin{array}{cc}
m_{1n}^2 \vec{V}'_{1n} \\
m_{2n}^2 \vec{V}'_{2n}
\end{array}\right)
 \ ,
\end{align}
where 
\begin{align}\notag
&O'_{11}=
\left(
\begin{array}{cc}
-\p_1^2-\p_2^2+4 q_2^2 x_2^2& -4q_1-2q_1x_1 \p_1 
\\ -2q_1+2q_1x_1 \p_1& 4q_1^2x_1^2+4q_2^2x_2^2-\p_2^2
\end{array}\right) \ , \\ 
&O'_{22}=
\left(
\begin{array}{cc}
-\p_1^2-\p_2^2+4 q_1^2 x_1^2&-4q_2-2q_2x_2\p_2
\\ -2q_2+2q_2x_2\p_2
&4q_1^2x_1^2+4q_2^2x_2^2-\p_1^2
\end{array}\right) \ ,
\end{align}
and 
\begin{align}
\vec{V}'_{1n}=\left(
\begin{array}{c}
\tilde{f}_{1n}\\ \tilde{g}_{1n}
\end{array}\right), \quad
\vec{V}'_{2n}=\left(
\begin{array}{c}
\tilde{f}_{2n}\\ \tilde{g}_{2n}
\end{array}\right) \ .
\end{align}
By solving these differential equations,
the eigenfunctions localized at the intersection point are obtained as
\begin{align} \label{tach}
\notag
\vec{V}'_{10}=
\left(
\begin{array}{c}
\tilde{f}_{10}(x_1,x_2,t)\\ \tilde{g}_{10}(x_1,x_2,t)
\end{array}\right)=
 e^{-q_1 x_1^2-q_2 x_2^2} C_{10}(t) 
\left(
\begin{array}{c}
1 \\ 1
\end{array} \right) \ ,
\\ 
\vec{V}'_{20}=
\left(
\begin{array}{c}
\tilde{f}_{20}(x_1,x_2,t)\\ \tilde{g}_{20}(x_1,x_2,t)
\end{array}\right)=
e^{-q_1 x_1^2-q_2 x_2^2} C_{20}(t) 
\left(
\begin{array}{c}
1 \\ 1
\end{array} \right) \ . 
\end{align}
The mass eigenvalues are obtained as
\begin{align} \notag
m_{10}^2&=2 (q_2-q_1) 
\sim \frac{1}{2\pi\alpha'} (\theta_2-\theta_1) \ , \\ 
m_{20}^2&=2 (q_1-q_2)
\sim \frac{1}{2\pi\alpha'} (\theta_1-\theta_2) \ .
\end{align}
There are massive and tachyonic modes, which coincide with the mass 
seen in the lowest mode of (\ref{stringmass}) in the small $q$ region.
Thus, we have obtained the correct lowest mode of the string 
mass spectrum in \cite{BDL}.

A brane system with a tachyon mode in open string theory is unstable and 
it rolls down to the stable vacuum by the condensation of the tachyon
mode\footnote{Stability of branes at angles is studied 
by considering the potential between 
two branes in \cite{potential}. 
} \cite{Sen}.
When we consider the condensation of the tachyon mode (\ref{tach}), 
we obtain
the configuration written as
\begin{align} \notag
&Y^9=\left(
\begin{array}{cc}
q_1 x_1&\tilde{f}_{10}\\
\tilde{f}_{10}&-q_1 x_1
\end{array}\right) , \quad
Y^8=\left(
\begin{array}{cc}
q_2 x_2&0\\
0&-q_2 x_2
\end{array}\right) \ , \\
&A_1=\left(
\begin{array}{cc}
0&i\tilde{g}_{10}\\
-i\tilde{g}_{10}&0
\end{array}\right) , \quad
A_2=0 \ ,
\end{align}
where we consider a case $q_1 >q_2$.
We cannot simultaneously diagonalize $Y^8$ and $Y^9$ by any gauge 
transformation, therefore, 
we do not find the simple recombination effect in the 
geometrical picture as was studied in \cite{Recombi}.
We will discuss more about this point in the next subsection.

Next, we search for the excited mode. We assume the form of the
eigenfunctions of the fluctuations as the product of exponential function
written as $e^{-q_1x_1^2-q_2x_2^2}$ 
and linear function of $x_a$. There are the following three eigenfunctions.
The first eigenfunctions are written as
\begin{align} \notag
\left(
\begin{array}{c}
\tilde{f}_{11}(x_1,x_2,t)\\ \tilde{g}_{11}(x_1,x_2,t)
\end{array}\right)=
x_2 e^{-q_1 x_1^2-q_2 x_2^2} C_{11}(t) 
\left(
\begin{array}{c}
1 \\ 1
\end{array} \right) \ ,
\\ 
\left(
\begin{array}{c}
\tilde{f}_{21}(x_1,x_2,t)\\ \tilde{g}_{21}(x_1,x_2,t)
\end{array}\right)=
x_1 e^{-q_1 x_1^2-q_2 x_2^2} C_{21}(t) 
\left(
\begin{array}{c}
1 \\ 1
\end{array} \right) \ .
\end{align}
The corresponding mass eigenvalues are obtained as
\begin{align} \notag
&m_{11}^2=2 (3 q_2 -q_1) \sim \frac{1}{2\pi\alpha'} (3\theta_2-\theta_1) \ , \\
&m_{21}^2=2 (3 q_1-q_2) \sim \frac{1}{2\pi\alpha'} (3\theta_1-\theta_1) \ .
\end{align}
The mass eigenvalues obtained here are found in the string mass spectrum in 
(\ref{stringmass}).
The second eigenfunctions are written as 
\begin{align} \notag
\left(
\begin{array}{c}
\tilde{f}_{12}(x_1,x_2,t)\\ \tilde{g}_{12}(x_1,x_2,t)
\end{array}\right)=
 x_1 e^{-q_1 x_1^2-q_2 x_2^2} C_{12}(t) 
\left(
\begin{array}{c}
1 \\ 1
\end{array} \right) \ ,
\\ 
\left(
\begin{array}{c}
\tilde{f}_{22}(x_1,x_2,t)\\ \tilde{g}_{22}(x_1,x_2,t)
\end{array}\right)=
 x_2 e^{-q_1 x_1^2-q_2 x_2^2} C_{22}(t) 
\left(
\begin{array}{c}
-1 \\ -1
\end{array} \right) \ ,
\end{align}
where the corresponding mass eigenvalues are common in both
eigenfunctions, that is, 
$C_{12}(t)=C_{22}(t)$. The mass eigenvalue is written as
\begin{align}
m_{i2}^2=2 (q_1+q_2) \sim \frac{1}{2\pi\alpha'} (\theta_1+\theta_2) \ ,
\end{align}
which is found in (\ref{stringmass}).

The third eigenfunctions written by the product of exponential function 
and linear function of $x_i$ are written as
\begin{align} \notag
\left(
\begin{array}{c}
\tilde{f}_{13}(x_1,x_2,t)\\ \tilde{g}_{13}(x_1,x_2,t)
\end{array}\right)=
q_1 x_1 e^{-q_1 x_1^2-q_2 x_2^2} C_{13}(t)
\left(
\begin{array}{c}
1 \\ 1
\end{array} \right) \ ,
\\ 
\left(
\begin{array}{c}
\tilde{f}_{23}(x_1,x_2,t)\\ \tilde{g}_{23}(x_1,x_2,t)
\end{array}\right)=
q_2 x_2 e^{-q_1 x_1^2-q_2 x_2^2} C_{23}(t)
\left(
\begin{array}{c}
1 \\ 1
\end{array} \right) \ , \label{fluc3}
\end{align}
where the squared of the mass eigenvalue is obtained as
\begin{align}
m_1^2=0 \ .
\end{align}
There is no massless mode in the string mass spectrum 
between intersecting branes at two different angles, except for some
fixed angles. 
This mode is considered as Nambu-Goldstone mode of broken U(2) symmetry
\footnote{For branes at one angle, 
it is discussed in \cite{EL}.}.
To see this explicitly,
let us consider a gauge transformation for the 
intersecting brane solution written as
\begin{align} 
Y^9=q_1 x_1 \sigma^3 ,
\quad
Y^8=q_2 x_2 \sigma^3 ,
\quad
A_1=A_2=0 \ .
\end{align}
When we consider the gauge transformation which is written as
\begin{align} \notag
&Y^8 \to \tilde{Y}^8 = U Y^8 U^{-1} \ , \quad
Y^9 \to \tilde{Y}^9 = U Y^9 U^{-1} \ , \\
&A_1 \to \tilde{A}_1 = U A_1 U^{-1} +i(\p_1 U) U^{-1} \ , \quad
A_2 \to \tilde{A}_2 = U A_2 U^{-1} +i(\p_2 U) U^{-1} \ ,
\end{align}
where 
\begin{align} \notag
&U(x_1,x_2)=e^{i \Lambda(x_1,x_2)} \ , \\
&\Lambda (x_1,x_2)=-\frac{C_3}{2} e^{-q_1x_1^2-q_2x_2^2} \sigma_2 \ ,
\end{align}
we obtain the configuration as follows:
\begin{align} 
\notag \label{artifact}
&\tilde{Y}^9=
q_1x_1(\sigma_3+C_3e^{-q_1x_1^2-q_2x_2^2} \sigma_1)
+{\cal O}(C_3^2) \ , \\ \notag
&\tilde{Y}^8=
q_2x_2(\sigma_3+C_3e^{-q_1x_1^2-q_2x_2^2} \sigma_1)
+{\cal O}(C_3^2) \ , \\ \notag
&\tilde{A}_1=-q_1x_1C_3e^{-q_1x_1^2-q_2x_2^2} \sigma_2 +{\cal O} (C_3^2)\ , \\
&\tilde{A}_2=-q_2x_2C_3e^{-q_1x_1^2-q_2x_2^2} \sigma_2 +{\cal O} (C_3^2)\ .
\end{align}
We find that the off-diagonal part in (\ref{artifact})
is equivalent with (\ref{fluc3}).
An off-diagonal massless mode which is not a Nambu-Goldstone mode
is considered in the next section.

Thus, we obtain the correct string mass of the lowest and first excited 
part of the spectrum in the approximation 
of small angles ${\cal O} (\theta_i)$.
We generalize these analyses to D3-branes intersecting at three angles,
which are written in appendix A.

\subsection{Tachyon condensation and recombination}

We focus on the tachyon mode (\ref{tach}) here.
By the condensation of the tachyon mode, 
we obtain the configuration written as
\begin{align} \notag
&Y^9=q_1 x_1 \sigma_3 +\tilde{f}_{10} \sigma_1 , \quad
Y^8=q_2 x_2 \sigma_3 \ , \\
&A_1=-\tilde{g}_{10} \sigma_2 , \quad
A_2=0 \ .
\end{align}
We can not simultaneously diagonalize $Y^8$ and $Y^9$
by any gauge transformation, but 
we can diagonalize $Y^8$ and $Y^9$ in some region simultaneously.
Let us look for the region where we can diagonalize $Y^8$ and $Y^9$
simultaneously.
In the region, 
\begin{align}
|x_1|\gg \frac{C_1}{q_1} \ ,
\end{align}
we can diagonalize $Y^8$ and $Y^9$ as
\begin{align}
Y^9 \sim q_1x_1 \sigma_3 \ , \quad
Y^8\sim q_2x_2 \sigma_3 \ .
\end{align}
Far away from the intersection point along $x_1$ direction, the brane
configuration remains unchanged, because  
the tachyon mode, which is due to the brane deformation, is 
localized near the intersection point.
Thus, this result is easily explained.
In the region $x_2 \sim 0$, we can also diagonalize $Y^8$ and $Y^9$ 
simultaneously as
\begin{align}
Y^9 \sim \sqrt{(q_1x_1)^2+\tilde{f}_{10}^2} \sigma_3 \ , \quad
Y^8 \sim 0 \ .
\end{align}
We find that 
the intersection point is resolved, and `recombination' occurs 
on the $x_1$-$Y^9$ plane near the intersection point. 
The other region can not be diagonalized and the D2-branes diffuse
in the $x_1x_2Y^8Y^9$ region near the intersection point. 
We will see the distribution of the
D2-brane charge in the next subsection.
Finally, let us remark a comment.
If $q_1 \gg q_2$, there are many tachyonic modes with masses $2(2 n
q_2+q_2-q_1)$ ($n=0,1,2,\cdots$).
In the limit $q_2 \to 0$, these infinite number of modes
will be summed up by the form written as 
$f \sim C e^{-q_1x_1^2}\sigma_3$, 
which is equivalent to the fluctuation mode found for branes at one angle
in \cite{Recombi}.

\subsection{D2-brane charge}

D2-brane charge density in terms of the D2-brane worldvolume effective action
is obtained by \cite{TR,Myers} as
\begin{align}
J_{0kl}=\frac{1}{12} 
\Tr (-i F^{ij} [Y^k,Y^l]-D_i Y^k D_j Y^l+D_i Y^l D_j Y^k) \ ,
\end{align}
where $i$ and $j$ are parameterizations of worldvolume directions.
D2-brane charge is already integrated along the direction
$Y^k$ and $Y^l$.

We consider the configuration written as
\begin{align} \label{susy}
\notag
&Y^9=q_1 x_1 \sigma_3 +\tilde{f}_{10} \sigma_1 \ , \quad
Y^8=q_2 x_2 \sigma_3 +\tilde{f}_{20} \sigma_1 \ , \\
&A_1=-\tilde{g}_{10} \sigma_2 \ , \quad
A_2=-\tilde{g}_{20} \sigma_2 \ .
\end{align}
If $q_1>q_2$, the combination ($\tilde{f}_{10},\tilde{g}_{10}$) 
corresponds to the tachyon mode
and ($\tilde{f}_{20},\tilde{g}_{20}$) corresponds to the massive mode.
We calculate the charge density of this configuration.
By the results,
\begin{align} \notag
-i\Tr  F_{12} [Y^8,Y^9]&= 8(C_2q_1x_1-C_1q_2x_2)^2 
e^{-2q_1x_1^2-2q_2x_2^2} \ , \\
-\Tr D_1 Y^8 D_2 Y^9&=8(C_2q_1x_1-C_1q_2x_2)^2
e^{-2q_1x_1^2-2q_2x_2^2} \ , \notag \\
\Tr D_1 Y^9 D_2 Y^8&= 2 q_1q_2-4(C_2^2q_1+C_1^2q_2)
e^{-2q_1x_1^2-2q_2x_2^2} \ ,
\end{align}
we obtain the D2-brane charge density as
\begin{align} \label{char}
J_{089}=\frac{1}{6} q_1q_2
+\left(\frac{2}{3}(C_2q_1x_1-C_1q_2x_2)^2-\frac{1}{3}C_2^2q_1-\frac{1}{3}
C_1^2q_2 \right) e^{-2q_1x_1^2-2q_2x_2^2} \ .
\end{align}
$C_1(t)$ is an exponentially growing function of $t$. 
We consider the situation where the tachyon mode condenses here, 
therefore, we take $C_2=0$ because $C_2$ is the coefficient 
of the massive mode.
D2-brane charge is rewritten as
\begin{align}
\frac{1}{6} q_1q_2
+\left(\frac{2}{3}(C_1q_2x_2)^2-\frac{1}{3}
C_1^2q_2 \right) e^{-2q_1x_1^2-2q_2x_2^2} \ .
\end{align}
The first term is background D2 charge.
The second term is induced by the tachyon mode and
localized near the intersection point.
Total D2-brane charge is obtained by the integration of $x_i$ as
\begin{align}
\int dx_1dx_2 e^{-2q_1x_1^2-2q_2x_2^2} 
\left(\frac{2}{3}(C_2q_1x_1-C_1q_2x_2)^2-\frac{1}{3}C_2^2q_1-\frac{1}{3}
C_1^2q_2 \right)=0 \ .
\end{align}
Thus, we confirm that D2-brane charge is still conserved
after tachyon mode has condensed.

\section{Supersymmetry}

In this section,  
we consider the case of equal intersection angles $q_1=q_2\equiv q$ here.
Now the mode (\ref{tach}) becomes massless and the corresponding
eigenfunctions are written as
\begin{align} \label{sudef}
\notag
\vec{V}'_{10}=
\left(
\begin{array}{c}
\tilde{f}_{10}(x_1,x_2,t)\\ \tilde{g}_{10}(x_1,x_2,t)
\end{array}\right)=
 e^{-q r^2} C_{10}
\left(
\begin{array}{c}
1 \\ 1
\end{array} \right) \ , 
\\ 
\vec{V}'_{20}=
\left(
\begin{array}{c}
\tilde{f}_{20}(x_1,x_2,t)\\ \tilde{g}_{20}(x_1,x_2,t)
\end{array}\right)=
 e^{-q r^2} C_{20}
\left(
\begin{array}{c}
1 \\ 1
\end{array} \right) \ ,
\end{align}
where $r^2 \equiv x_1^2+x_2^2$. $C_{10}$ and $C_{20}$ 
are some numerical constant.
It is known that two intersecting D-branes at equal two angles preserve 
$1/4$ of the supersymmetries \footnote{Black hole entropy in this system
is studied in \cite{Costa}.}.
 Therefore,
let us check the supersymmetry of the configuration written as
\begin{align} \label{susycon}
\notag
&Y^9=q_1 x_1 \sigma_3 +\tilde{f}_{10} \sigma_1 \ , \quad
Y^8=q_2 x_2 \sigma_3 +\tilde{f}_{20} \sigma_1 \ , \\
&A_1=\tilde{g}_{10} \sigma_2 \ , \quad
A_2=\tilde{g}_{20} \sigma_2 \ .
\end{align}

The supersymmetric variation of gaugino is written as
\begin{align}
\delta \psi =F_{\mu\nu} \Gamma^{\mu\nu} \epsilon \ .
\end{align}
The configuration (\ref{susycon}) satisfies the following BPS conditions
up to the quadratic order in the fluctuations as
\begin{align} \notag
F_{12}+ i [Y^8,Y^9]&=0 \ , \\ \notag
D_1 Y^8+D_2 Y^9&=\sigma^3 {\cal O}(C^2) \ , \\
D_1 Y^9-D_2 Y^8&=\sigma^3 {\cal O}(C^2) \ ,
\end{align}
and therefore, we obtain 
\begin{align}
\delta \psi = {\cal O} (C^2) \epsilon \ .
\end{align}
Thus, this configuration is supersymmetric up to the quadratic order in the
fluctuations. 
The configuration with the supersymmetry up to quadratic order and
where all supersymmetries are broken beyond this order is studied in
\cite{Taylor1}.

We can diagonalize $Y^8$ and $Y^9$ simultaneously
in the region $x_1 \sim x_2 \sim 0$ as
\begin{align}
Y^9\sim \tilde{f}_{10} \sigma_3 \ , \quad
Y^8\sim \tilde{f}_{20} \sigma_3 \ .
\end{align}
Thus, the intersection point is resolved, which was also seen in 
the previous section.
Note that in the previous section, 
only one mode, which is tachyonic, condenses. On the other hand,
we consider the deformation by two massless modes here.
Finally,
we calculate the distribution of the D2-brane charge.
D2-brane charge is written in (\ref{char}) as
\begin{align}
J_{089}=\frac{1}{6} q^2
+\left(\frac{2}{3}(C_2x_1-C_1x_2)^2q^2-\frac{1}{3}C_2^2q-\frac{1}{3}
C_1^2q \right) e^{-2qr^2} \ .
\end{align}
Total charge is obtained as 
\begin{align}\notag
\int dx_1dx_2&\left( \frac{1}{6} q^2
+\left(\frac{2}{3}(C_2x_1-C_1x_2)^2q^2-\frac{1}{3}C_2^2q-\frac{1}{3}
C_1^2q \right) e^{-2qr^2} \right) \\
&=\frac{q^2}{6} \cdot 
\text{(area of $x_1$-$x_2$ plane)} \ .
\end{align}
Thus, total D2-brane charge is again conserved.

\section{Higher order corrections of $F$}

We consider the effect of the higher order corrections of $F$.
The $F^4$ terms are the first nontrivial contribution to the 
fluctuation analysis, therefore, we consider the mass spectra and 
eigenfunctions including $F^4$ terms.
The symmetrized traced Lagrangian in \cite{tse} is written as
\begin{align}
L={\rm Str} \sqrt{- \det (\eta_{\mu\nu} +2 \pi\alpha' F_{\mu\nu})} \ ,
\end{align}
and $F^4$ terms are obtained by the expansion of this action as
\begin{align}
L={\rm Str} (2\pi\alpha')^2
\left(\frac{1}{8} F_{\mu\nu} F_{\nu\lambda} F_{\lambda\sigma} F_{\sigma\mu}
-\frac{1}{32} F_{\mu\nu} F_{\nu\mu} F_{\lambda\sigma} F_{\sigma\lambda}
\right) \ .
\end{align}
By taking the T-duality along $8,9$ directions and considering the
classical solutions (\ref{sol1}) and fluctuations (\ref{flu1}),
we obtain quadratic parts in the fluctuations as
\begin{align} \notag
L=&-
Q_1\left((\p_0f_1)^2+(\p_0g_1)^2\right)
-Q_2\left((\p_0f_2)^2+(\p_0g_2)^2\right) \\ \notag
&+Q_3
\left(
(\p_1f_1+2g_1q_1x_1)^2-4 q_1f_1g_1\right)
+Q_4\left(
(\p_2f_2+2g_2q_2x_2)^2-4 q_2f_2g_2\right)\\ \notag
&+Q_5\left(
(\p_1f_2+2g_1q_2x_2)^2+(\p_2f_1+2g_2q_1x_1)^2
+4(q_1x_1f_2-q_2x_2f_1)^2+(\p_1g_2-\p_2g_1)^2
\right)\ , \\ \notag
&Q_1\equiv(1-\frac{1}{6} \frac{q_1^2}{(2\pi\alpha')^2})
(1+\frac{1}{6}\frac{q_2^2}{(2\pi\alpha')^2}) \ , \quad 
Q_2\equiv(1+\frac{1}{6} \frac{q_1^2}{(2\pi\alpha')^2})
(1-\frac{1}{6}\frac{q_2^2}{(2\pi\alpha')^2}) \ , \\ \notag
&Q_3\equiv(1-\frac{1}{2}\frac{q_1^2}{(2\pi\alpha')^2})
(1+\frac{1}{6}\frac{q_1^2}{(2\pi\alpha')^2}) \ , \quad 
Q_4\equiv(1+\frac{1}{6}\frac{q_1^2}{(2\pi\alpha')^2})
(1-\frac{1}{2}\frac{q_1^2}{(2\pi\alpha')^2}) \ , \\
&Q_5\equiv(1-\frac{1}{6}\frac{q_1^2}{(2\pi\alpha')^2})
(1-\frac{1}{6}\frac{q_1^2}{(2\pi\alpha')^2})  \ .
\end{align}
By solving the equations of motions for the fluctuations,
we obtain the eigenfunctions of the lowest mode as
\begin{align} \notag
\vec{V}'_{10}=
\left(
\begin{array}{c}
\tilde{f}_{10}(x_1,x_2,t)\\ \tilde{g}_{10}(x_1,x_2,t)
\end{array}\right)=
n_{10} e^{-q_1 x_1^2-q_2 x_2^2} C_{10}(t) 
\left(
\begin{array}{c}
1 \\ 1
\end{array} \right) \ , 
\\ 
\vec{V}'_{20}=
\left(
\begin{array}{c}
\tilde{f}_{20}(x_1,x_2,t)\\ \tilde{g}_{20}(x_1,x_2,t)
\end{array}\right)=
n_{20} e^{-q_1 x_1^2-q_2 x_2^2} C_{20}(t) 
\left(
\begin{array}{c}
1 \\ 1
\end{array} \right) \ . 
\end{align}
The squared of the mass eigenvalue is obtained as
\begin{align}
m_{i0}^2=\pm (2 (q_1-q_2)-\frac{2}{3(2\pi\alpha')^2}(q_1^3-q_2^3))
=\pm \frac{(\theta_1-\theta_2)}{2\pi\alpha'}+{\cal O}(\theta_i^5) \ .
\end{align}
We obtain the correct mass eigenvalue of tachyon mode at order $\theta^3$. 
The eigenfunction of tachyon mode is the product of Gaussian functions
of $x_1$ and $x_2$. In the analysis \cite{Nagaoka1}, the eigenfunctions
of tachyon mode remain Gaussian function at order $F^6$, therefore 
we expect that the eigenfunctions remain the product of Gaussian functions
of $x_1$ and $x_2$ even in two angle's case in the action including
the $F^6$ terms, and further, more higher order $F$ terms.

\section{Conclusion and discussion}

We have considered the recombinations of D-branes intersecting at
more than one angle using SU(2) super Yang-Mills theory. 
There are tachyon modes in the off-diagonal fluctuations and 
the condensation of this mode triggers the recombinations.
On the other hand, in supersymmetric intersecting brane systems,
there are two kinds of nontrivial massless deformations, 
the deformations into calibration geometry
and the recombinations by the condensation of the off-diagonal
fluctuation modes.
The former is obtained by solving a minimal surface problem of membranes and 
described by U(1) DBI actions.
On the other hand, to describe the latter phenomenon,
we need to know the full knowledge of the non-abelian Born-Infeld action. 
In particular configurations, 
we can analyze this phenomenon even in Yang-Mills theory.
We have studied the condensation of the tachyon mode and we have found that 
there are D2-branes distributed in the bulk near the intersection point 
after tachyon mode had condensed.
Tachyon condensation diffuses D2-brane charge
at the intersection point at first stage,
and after that, the localization might be relaxed and 
recombined D2-branes, which preserve $1/4$ of the supersymmetries, 
would emerge as a final state.
In the case that two intersection angles are equal,
the lowest off-diagonal fluctuation mode is massless.
The configuration deformed by this mode is supersymmetric 
up to the quadratic fluctuations, and 
the intersection point is resolved.
This is an interesting example of non-abelian embeddings. 
In this case, abelian mode which governs the calibration geometry
and non-abelian mode which governs the non-abelian embeddings.
The region that Yang-Mills analysis is appropriate is also 
up to the quadratic order in the fluctuations. 
In string theory, or NBI action, this supersymmetric non-abelian
embedding might be valid beyond the quadratic order in the fluctuations.
To discuss this point more deeply,
it might be interesting to consider the higher order fluctuations.
$F^4$ corrections are also studied and the higher order corrections 
of $\theta$ are obtained.
It is straightforward to discuss the recombination of D$p$-branes 
intersecting at more angles. 
The extension to three intersection angles is studied
in appendix A. 
In future direction,
it might be interesting to study the Higgs mechanism of the Standard model 
from this direction.

\section*{Acknowledgements}

I would like to thank M. Kato, M. Naka and especially K. Hashimoto for
quite useful discussions and comments. I would also like to thank
K. Hashimoto for carefully reading this manuscript.

\appendix

\section{Recombination of D3-branes intersecting at three angles}

D3-branes effective action is written as
\begin{align}
S= -T \ \Tr \int d^2x dt \left[ (D_a Y^i)^2 +\frac{1}{2}F_{ab}^2 
-[Y^i,Y^j]^2 \right] \ .
\end{align}
where the indices $a,b=0,1,2,3$ and  
$i,j=4,\cdots,9$.
Two D3-branes are embedded in 6 dimensions and 
do not extend in other dimensions.
The embedded directions are chosen as $1,2,3,7,8$ and $9$ and
others as $4,5$ and $6$.
A classical solution we consider here is written as
\begin{align}
Y^9=q_1 x_1 \sigma^3 ,
\quad
Y^8=q_2 x_2 \sigma^3 ,
\quad
Y^7=q_3 x_3 \sigma^3 ,
\quad
A_a=0 \ ,
\end{align}
which describes the two intersecting D3-branes.

Let us turn on the off-diagonal fluctuations as
\begin{align} \notag
&Y^9=q_1 x_1 \sigma^3 +f_1(x_a) \sigma^1-\bar{f}_1 (x_a)\sigma^2 , \quad
Y^8=q_2 x_2 \sigma^3 +f_2(x_a) \sigma^1 -\bar{f}_2 (x_a) \sigma^2 \ , \\
&A_1=g_1(x_a) \sigma^1-\bar{g}_1(x_a)\sigma^2 , \quad
A_2=g_2(x_a) \sigma^1-\bar{g}_2(x_a)\sigma^2 . 
\end{align}

\begin{figure}[t]
\begin{center}
\begin{minipage}{16cm}
\begin{center}
\includegraphics[width=14cm]{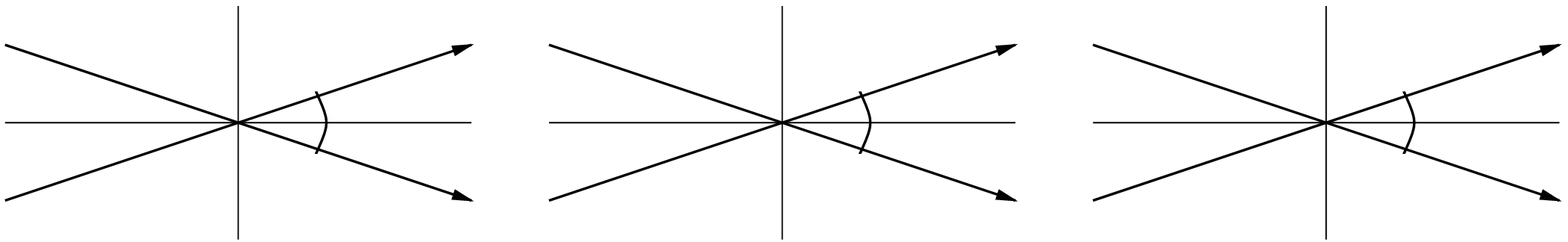} 
\put(-280,35){$x_1$}
\put(-140,35){$x_2$} 
\put(0,35){$x_3$}
\put(-55,60){$Y^7$}
\put(-195,60){$Y^8$} 
\put(-335,60){$Y^9$}
\put(-300,35){$\theta_1$}
\put(-160,35){$\theta_2$}
\put(-20,35){$\theta_3$}
\caption{Intersecting D3-branes} 
\end{center}
\end{minipage}
\end{center}
\end{figure}

The Lagrangian quadratic in the fluctuations is calculated as
\begin{align}\notag
L=\sum_{i,j=1,2,3}&
\Big(-(\p_0 f_i)^2 -
4q_if_i\bar{g}_i+
(\p_i f_j+2\bar{g}_iq_jx_j)^2
-(\p_0 \bar{g}_i)^2
+(\p_i\bar{g}_j-\p_j\bar{g}_i)^2 \\ 
&+4 (q_i x_i f_j-q_j x_j f_i)^2 
+(f_\alpha \to \bar{f},\bar{g}_\alpha \to g ) \Big)
\ .
\end{align}
The combinations ($f_\alpha , \bar{g}_\alpha $) and 
($\bar{f}_\alpha , g_\alpha $) are decoupled each other in the quadratic
fluctuations, therefore we neglect ($\bar{f}_\alpha , g_\alpha$).
From now on, we denote $\bar{g}$ as $g$.

By solving the equations of motion for the fluctuations, we obtain 
the eigenfunction of the lowest mode of the fluctuations as
\begin{align} \notag
\left(
\begin{array}{c}
\tilde{f}_{10}(x_1,x_2,t)\\ \tilde{g}_{10}(x_1,x_2,t)
\end{array}\right)=
\sum _{i=1,2,3}
n_1 e^{-q_i x_i^2} C_{10}(t) 
\left(
\begin{array}{c}
1 \\ 1
\end{array} \right) \ , 
\\ 
\left(
\begin{array}{c}
\tilde{f}_{20}(x_1,x_2,t)\\ \tilde{g}_{20}(x_1,x_2,t)
\end{array}\right)=
\sum_{i=1,2,3}
n_2 e^{-q_i x_i^2} C_{20}(t) 
\left(
\begin{array}{c}
1 \\ 1
\end{array} \right) \notag \ , \\
\left(
\begin{array}{c}
\tilde{f}_{30}(x_1,x_2,t)\\ \tilde{g}_{30}(x_1,x_2,t)
\end{array}\right)=
\sum _{i=1,2,3}
n_3 e^{-q_i x_i^2} C_{30}(t) 
\left(
\begin{array}{c}
1 \\ 1
\end{array} \right) \ .
\end{align}
The lowest mode remains the product of Gaussian functions of $x_i$.
The squared of the mass eigenvalue is obtained as\footnote{The 
mass spectra of constant field strength configuration in Yang-Mills 
theory on $T^4$ are studied in \cite{VanBaal} and the generalizations 
on $T^{2n}(n=1,2,\cdots)$ are studied in \cite{Troost}.}
\begin{align} \notag
m_{10}^2=\pm 2 (-q_1+q_2+q_3) \sim \frac{1}{2\pi\alpha'} 
(-\theta_1+\theta_2+\theta_3) \ ,
\\ \notag
m_{20}^2=\pm 2 (q_1-q_2+q_3) \sim \frac{1}{2\pi\alpha'} 
(\theta_1-\theta_2+\theta_3) \ , \\
m_{30}^2=\pm 2 (q_1+q_2-q_3) \sim \frac{1}{2\pi\alpha'} 
(\theta_1+\theta_2-\theta_3) \ .
\end{align}
This result is consistent with the worldsheet analysis in \cite{BDL}.
The parameter region of the intersection angles which corresponds to the 
 tachyonic configuration is 
considered in \cite{Rabadan}.

\newcommand{\J}[4]{{\sl #1} {\bf #2} (#3) #4}
\newcommand{\andJ}[3]{{\bf #1} (#2) #3}
\newcommand{\AP}{Ann.\ Phys.\ (N.Y.)}
\newcommand{\MPL}{Mod.\ Phys.\ Lett.}
\newcommand{\NP}{Nucl.\ Phys.}
\newcommand{\PL}{Phys.\ Lett.}
\newcommand{\PR}{ Phys.\ Rev.}
\newcommand{\PRL}{Phys.\ Rev.\ Lett.}
\newcommand{\PTP}{Prog.\ Theor.\ Phys.}
\newcommand{\hep}[1]{{\tt hep-th/{#1}}}

\end{document}